\documentclass[prb,preprint,superscriptaddress,citeautoscript,preprintnumbers,showpacs]{revtex4}

\usepackage[dvips]{graphicx}
\usepackage{epsfig}
\usepackage{dcolumn}

\usepackage{amsmath}

\newcommand*{\Et}{\mathcal{E}}

\begin{document}

\title{\emph{Ab initio} supercell calculations on nitrogen-vacancy
center in diamond: its electronic structure and hyperfine tensors}

\author{Adam Gali} \affiliation{Department of Atomic Physics, Budapest
  University of Technology and Economics, Budafoki \'ut 8., H-1111, 
  Budapest, Hungary}
  \affiliation{Department of Physics and School of Engineering and 
Applied Sciences, Harvard University, Cambridge, Massachusetts 02138, USA}

\author{Maria Fyta} \affiliation{Department of Physics and School of 
Engineering and Applied Sciences, Harvard University, Cambridge, Massachusetts 02138, USA}

\author{Efthimios Kaxiras} \affiliation{Department of Physics and School of
  Engineering and Applied Sciences, Harvard University, Cambridge, Massachusetts 02138, USA}

\begin{abstract}
  The nitrogen-vacancy center in diamond is a promising candidate for
  realizing the spin qubits concept in quantum information. Even
  though this defect is known for a long time, its electronic
  structure and other properties have not yet been explored in detail.
  We study the properties of the nitrogen-vacancy center in diamond
  through density functional theory within the local spin density
  approximation, using supercell calculations. While this theory is
  strictly applicable for ground state properties, we are able to give
  an estimate for the energy sequence of the excited states of this
  defect. We also calculate the hyperfine tensors in the ground state.
  The results clearly show that: (i) the spin density and the
  appropriate hyperfine constants are spread along a plane and
  unevenly distributed around the core of the defect; (ii) the
  measurable hyperfine constants can be found within about 7~\AA\ from
  the vacancy site. These results have important implications on the
  decoherence of the electron spin which is crucial in realizing the
  spin qubits in diamond.
\end{abstract}

\maketitle

\section{Introduction}
The nitrogen-vacancy (NV) center in diamond has attracted much
attention in recent years, because it has been shown to give rise to a single
optically active level within the diamond band gap~\cite{Gruber97,Drabenstedt99} ,
and as such provides an interesting candidate for a qubit for
quantum computing applications
\cite{Wrachtrup01,Jelezko02,Jelezko04-1,Jelezko04-2,Epstein05,Hanson06PRL}.
Besides providing a single photon source for quantum cryptography
\cite{Brouri00,Beveratos02}, the NV center is also a promising
candidate as an optically coupled quantum register for scalable
quantum information processing, such as quantum communication
\cite{Childress06-PRL} and distributed quantum computation
\cite{Jiang07}. In addition, it has been recently demonstrated that
proximal nuclear spins can be coherently controlled via hyperfine
interaction \cite{Childress06} and used as a basis for quantum memory with an
extremely long coherence time \cite{Childress07}. Therefore, knowing
the electron-nucleus hyperfine interaction and its position-dependence
is essential to analyze and optimize coherent control of proximal
nuclear spins \cite{Cappellaro07}.

Experimentally, the hyperfine constants of the closest atoms near the
vacancy are known from electron paramagnetic resonance (EPR) and
electron-nuclear double resonance (ENDOR) studies
\cite{Loubser77,He93}.  The hyperfine interaction between $^{13}$C
isotopes farther from the vacancy contributes to the coherent
electron-nuclear spin states in the measurements
\cite{Childress06,Childress07}. \emph{Ab initio} supercell
calculations can be a very useful tool for determining the hyperfine
tensors of a defect. For instance, such calculations have been used to
identify the basic vacancy defects in silicon carbide by comparing the
measured and calculated hyperfine constants
\cite{UmedaPRB05,UmedaPRL06,SonPRL06,UmedaPRB07}. In the present paper we
focus on the calculation of the full hyperfine tensor of
the NV center in diamond which is of very high importance for qubit 
applications. Previous theoretical work has reported a calculation of
the hyperfine constants of the NV center in a small 64-atom supercell
\cite{Luszczek04}, but that work determined only the Fermi-contact term rather than the 
full hyperfine tensor.  We will show in
the Results and Discussion section that the conclusions based on that
earlier analysis were adversely affected by the small unit cell size.
The larger supercell of 512 atoms employed here does not suffer from 
this limitation and provides a realistic picture for the defect properties,
which are in excellent agreement with experimental measurements.

The rest of this paper is organized as follows: Section II gives a general 
discussion of the electronic states of the NV center based on a
single-particle picture and the many-body states that can be
constructed from this basis. Section III describes the method of the
first-principles calculations we performed.  Section IV presents and
discusses our results concerning the atomic and electronic structure 
of the defect, as obtained from the {\em ab initio} calculations. 
In Section V we present a detailed discussion of the
calculated  hyperfine interactions.
Finally, we give our conclusions on the nature of this defect
in Section VI.

\section{The electronic states of the NV center in diamond}

The electronic structure of NV center in diamond has been discussed in
detail in a recent paper~\cite{Manson06}; we briefly review the
main points here. The NV center was found many years ago in diamond
\cite{duPreez65}. The concentration of NV centers can be enhanced in
N-contaminated diamond by irradiation and annealing
\cite{duPreez65,Davies76}. The model of the NV center consists of a
substitutional nitrogen atom adjacent to a vacancy in diamond
\cite{duPreez65,Davies76,Loubser77,Collins83}. The NV center has a
strong optical transition with a zero phonon line (ZPL) at 1.945~eV
(637~nm) accompanied by a vibronic band at higher energy in absorption
and lower energy in emission. Detailed analysis of the ZPL revealed
that the center has trigonal, C$_{3v}$ symmetry \cite{Davies76}.
Later, an optically induced EPR center was found in diamond which
correlated with the NV center \cite{Loubser77}. The EPR center showed
trigonal symmetry with a spin polarized triplet state (S=1). Since the
nitrogen atom has five valence electrons and the S=1 state implies
even number of electrons, the NV defect must be charged in the EPR
measurement. It was assumed that the NV defect is negatively charged
and the extra electron may be donated from isolated substitional
nitrogen defects \cite{Loubser77}. In a recent measurement, the
coupling between the NV center and the nitrogen substitional has been
indeed detected \cite{Hanson06}. Loubser and van Wyk~\cite{Loubser77} measured
the NV EPR signal just under the optical excitation, based on which they proposed
that the spin polarization arises from a singlet electronic system
with inter-system crossing to a spin level of a metastable triplet.
Redman and co-workers~\cite{Redman91} detected the NV center in the
dark even at 100~K by EPR, from which they concluded that the S=1 state is the
ground state of the NV center. Hole burning~\cite{Reddy87}, 
optically detected magnetic resonance~\cite{Oort88}
and Raman heterodyne measurements~\cite{Manson92} also showed the S=1
state to be the ground state of the NV center.

A group theory analysis based on a single-particle picture can be very useful
in understanding the nature of the defect states and the
possible optical transitions between them.
While the number of electrons in the NV center has been disputed in
the literature~\cite{Goss97,Lenef97} a previous \emph{ab initio}
calculation clearly supported the negatively charged NV defect
\cite{Goss96} as was originally proposed by Loubser and van Wyk
\cite{Loubser77}. We will also show in the Results and Discussion
section that the NV center should be negatively charged. 

\begin{figure}[h]
  \centering
  \includegraphics[keepaspectratio,width=8.5cm]{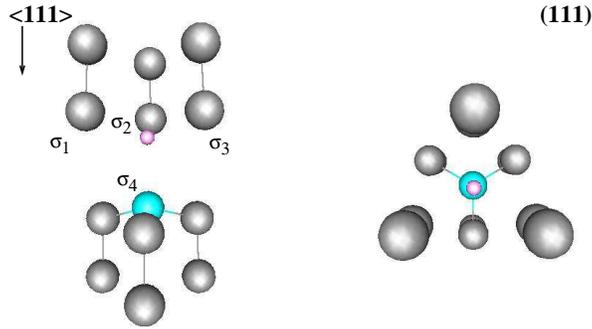}
  \caption{(Color online) The NV center viewed in perspective (left)
    and along the $<$111$>$ direction (right); this direction is the C$_3$
    symmetry axis of the C$_{3v}$ symmetry group of the defect. 
    The vacant site is indicated by a small pink circle and the
    neighboring C and N atoms by grey and
    cyan balls, respectively. The first neighbor atoms around the
    vacant site have $sp^3$ dangling bonds pointing toward the vacant
    site, which
    are labeled $\sigma_i~(i=1,4)$, as in the scheme used in the group theory analysis.}
  \label{fig:states}
  \centering 
\end{figure}

In the NV defect, three carbon atoms have $sp^3$ dangling bonds near the vacancy
and three back bonds each to the lattice, while the nitrogen atom has
also three back bonds and one dangling bond pointing to the vacant
site. Since nitrogen has five valence electrons the negatively charged
NV defect has altogether six electrons around the vacant site. The
structure of the NV defect, including the definition of the symmetry
$<$111$>$ axis, is depicted in Fig.~\ref{fig:states}.

\begin{figure}[t]
  \centering
 \includegraphics[keepaspectratio,width=8cm]{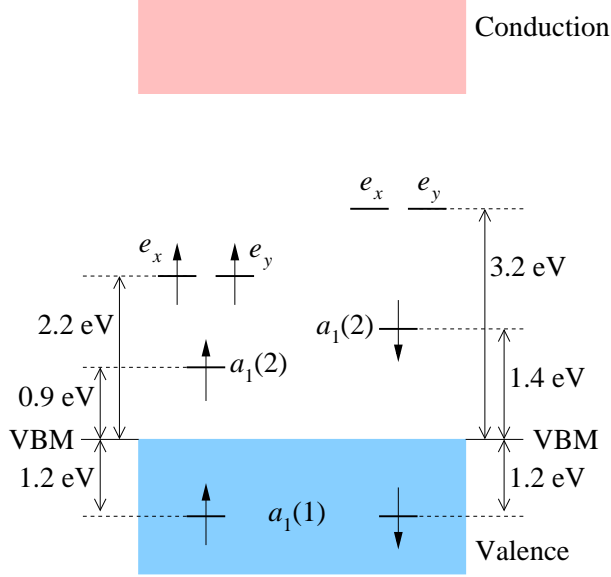}
  \caption{(Color online) The calculated spin-resolved single-electron
    levels with respect to the valence band maximum (VBM) in the
    ground state of the NV defect.  Valence and conduction bands of
    the host crystal are shown as blue and pink shaded regions,
    respectively.  The levels are labeled as in Eq.
    (\ref{eq:orbitals}) and their occupation is given for a negatively
    charged defect (a total of 6 electrons).}
  \label{fig:levels}
  \centering 
\end{figure}

The group theory analysis of the six electron model has been worked
out previously for this defect~\cite{Lenef96}. We summarize the
results using our notation and conventions: Since it is known that
the carbon and nitrogen atoms relax outward from the vacancy
\cite{Goss96,Luszczek04}, we assume that the overlap between the
dangling bonds ($\sigma _{1-4}$) is negligible, that is, $\sigma _i \sigma
_j = \delta _{ij}$. $\sigma _{1-4}$ are transformed under the
operation of C$_{3v}$ point group forming the following orthonormal
states: 
\begin{eqnarray}
\centering
\label{eq:orbitals}
a_1(1)&:&  \phi _1 = \sqrt{1-\alpha ^2}\; \sigma _4 -
\frac{\alpha}{\sqrt{3}}\; ( \sigma _1 + \sigma _2 + \sigma _3) \nonumber \\
a_1(2)&:& \phi _2 = \alpha \; \sigma _4 + 
\sqrt{\frac{1-\alpha ^2}{3}}\; ( \sigma _1 + \sigma _2 + \sigma _3) \\
e_x&:& \phi _3 = \frac{1}{\sqrt{6}}\; ( 2 \sigma _1 - \sigma _2 - \sigma
_3) \nonumber \\
e_y&:&  \phi _4 = \frac{1}{\sqrt{2}}\; ( \sigma _2 - \sigma _3) \nonumber
\end{eqnarray}
where $0\leq \alpha \leq 1$is a parameter that determines the extent to which the nitrogen
dangling bond is mixed in the $\phi _1$ and $\phi _2$ defect
states. There are two fully symmetric one-electron states ($a_1$)
and one doubly degenerate $e$ state, with a total occupation of 6
electrons. We note here that the dangling bond of nitrogen is
\emph{not} mixed in the $e$ state but only in the $a_1$ states. It
was found by Goss \emph{et al.}~\cite{Goss96} using \emph{ab initio} molecular cluster
calculations that the two $a_1$ states are lower in energy than the
$e$ state. As a consequence, four electrons occupy the $a_1$ states
and two electrons remain for the $e$ state. Our calculated
one-electron levels obtained by \emph{ab initio} supercell
calculations are shown in Fig.~\ref{fig:levels}.  
As can be seen from this analysis, the natural choice is to put the
two remaining electrons in the $e$ level forming an S=1 state (by
analogy to Hund's rule for the $p$-orbitals of the isolated group IV
elements in the Periodic Table). 

In the C$_{3v}$ point group the total wavefunction has $^3A _2$
symmetry with S=1. In our special case we choose $M_\text{S} = 1$, so
both electrons are spin-up electrons in the $e$ level. The C$_{3v}$
symmetry can also be maintained by other occupations of the states.
Putting two electrons into four possible quantum states of the
degenerate $e$ level, we end up with six possible multiplets
(including the degeneracy): $^3A _2$, $^1A _1$, and $^1E$.  By taking
the $^3A _2$ to be the ground state of the defect (as experiments
indicate) there is no allowed optical transition to first order, since
the spin state cannot be changed in a PL process. The $\phi _1$ level
is relatively deep in the valence band, so to a good approximation we
can assume that it does not contribute to the excitation process.
However, the $\phi _2$ level in the gap is not very far from the $e$
level. If one electron is excited from $\phi _2$ into the $e$ level
($\phi _3$ or $\phi _4$) then either a $^3E$ or a $^1E$ multiplet is
obtained. If both electrons are excited from $\phi _2$ to $\phi _3$
and $\phi _4$ then a fully symmetric $^1A _1$ state is obtained. The
only allowed transition is $^3A _2 \rightarrow ^3E$ to first order.
The electronic configurations of these states are explained in
Table~\ref{tab:conf}. Most of these states were discussed in
Ref.~\onlinecite{Lenef96}.

\begin{table}
\caption{\label{tab:conf} The electronic configurations and the
  possible total wave functions with C$_{3v}$ symmetry. For
  simplicity we abbreviate $\sigma _2 \rightarrow 2$, etc.\ in the
  last column. Overbar in a wavefunction means spin-down electrons, the
  rest are spin-up electrons. We assume that $a_1(1) : \sigma _1$ is
  fully occupied, so we do not show that part of the wave function
  here. In the second and third columns we give the symmetry of the
  total wavefunction ($\Gamma$) and its spin projection ($M_\text{S}$),
  respectively. In the case of doubly degenerate representations ($E$
  states) we designate which transforms as x or y in the last
  column.}
\begin{ruledtabular}
\begin{tabular}{ccccc}
Configuration  &  $\Gamma$  & $M_\text{S}$ & (x,y)  & Wave function \\ \hline

$a_1^2(2)e^2$  &  $^3A _2$  &   1    &       & $|2\bar{2}34\rangle$ \\

               &            &   0    &       & $\frac{1}{\sqrt{2}}[
                                              |2\bar{2}3\bar{4}\rangle +
                                              |2\bar{2}\bar{3}4\rangle]$ \\      
               &            &  -1    &       & 
                                       $|2\bar{2}\bar{3}\bar{4}\rangle$ \\
               &  $^1A _1$  &   0    &       &$\frac{1}{\sqrt{2}}[ 
                                              |2\bar{2}3\bar{3}\rangle +
                                              |2\bar{2}4\bar{4}\rangle]$ \\
               &  $^1E$     &   0    &  x    &$\frac{1}{\sqrt{2}}[ 
                                              |2\bar{2}3\bar{3}\rangle -
                                              |2\bar{2}4\bar{4}\rangle]$ \\
               &            &        &  y    & $\frac{1}{\sqrt{2}}[
                                              |2\bar{2}3\bar{4}\rangle -
                                              |2\bar{2}\bar{3}4\rangle]$ \\
$a_1^1(2)e^3$  &  $^3E$     &   1    &  x    &$|234\bar{4}\rangle$     \\
               &            &        &  y    &$|23\bar{3}4\rangle$     \\
               &            &   0    &  x    &$\frac{1}{\sqrt{2}}[
                                               |\bar{2}34\bar{4}\rangle+
                                               |2\bar{3}4\bar{4}\rangle]$ \\
               &            &        &  y    &$\frac{1}{\sqrt{2}}[
                                               |\bar{2}3\bar{3}4\rangle+
                                               |23\bar{3}\bar{4}\rangle]$ \\
               &            &  -1    &  x
               &$|\bar{2}\bar{3}4\bar{4}\rangle$ \\
               &            &        &  y
               &$|\bar{2}3\bar{3}\bar{4}\rangle$ \\
               &  $^1E$     &   0    &  x    &$\frac{1}{\sqrt{2}}[
                                               |\bar{2}34\bar{4}\rangle-
                                               |2\bar{3}4\bar{4}\rangle]$ \\            
               &            &        &  y    &$\frac{1}{\sqrt{2}}[
                                               |\bar{2}3\bar{3}4\rangle-
                                               |23\bar{3}\bar{4}\rangle]$ \\   
$a_1^0(2)e^4$  &  $^1A _1$  &   0    &       &$|3\bar{3}4\bar{4}\rangle$     \\         
\end{tabular}
\end{ruledtabular}
\end{table}

The $M_\text{S}=\pm 1$ triplet states can be described
by a single Slater-determinant. However, the singlet states (except
for the last $^1A _1$ state) can be described by a linear combination of
two Slater-determinants. The two singlet single-Slater-determinant
states of the $a_1^2(2)e^2$ configuration are:
\begin{subequations}
\begin{equation}
\label{eq:multi1a}
|\sigma_2\bar{\sigma_2}\sigma_3\bar{\sigma_4}\rangle =
 \frac{1}{\sqrt{2}}[^1E(0,y) + ^3A_2(0)]
\end{equation}
\begin{equation}
\label{eq:multi1b}
|\sigma_2\bar{\sigma_2}\sigma_3\bar{\sigma_3}\rangle = \frac{1}{\sqrt{2}}[^1E(0,x) + ^1A_1(0)]
\end{equation} 
\end{subequations} 
where $^1E(0,\lambda)$ is the multideterminant wavefunction of
the singlet $E$ state of $M_\text{S}$=0 at $\lambda$-row ($\lambda=$ x or y) in
Table~\ref{tab:conf}. The singlet
single-Slater-determinant state of the $a_1^1(2)e^3$ configuration is:
\begin{equation}
\label{eq:multi2}
|\bar{\sigma_2}\sigma_3\sigma_4\bar{\sigma_4}\rangle = \frac{1}{\sqrt{2}}[^3E(0,x) + ^1E(0,x)]
\end{equation}  

In summary, the following many-body states must be considered:
$^3A_2$, $^1A_1$, $^1E$, $^3E$, $^1E$, and $^1A_1$. The two triplet
states are orthogonal to each other while the two $^1A_1$ and the two
$^1E$ states theoretically can be mixed with each other. This will be
discussed below. Our computational method described in the next
section cannot take the spin-orbit and spin-spin interaction into
account. From the energetic point of view those effects are marginal
(within few meV) but they could have important consequences on the
possible optical transitions and the spin state of the NV center
\cite{Manson06}.

\section{Computational Method}

We use density functional theory
with the local spin density approximation (DFT-LSDA) of Ceperley-Alder
\cite{C-A80} as parameterized by Perdew and Zunger \cite{Perdew81}. We
employed three different codes and somewhat different methodologies
to carry out the calculations. The geometry of the defect was
optimized with the {\sc VASP} code~\cite{Kresse94,Kresse96} and the
{\sc  SIESTA} code~\cite{Siesta}. The latter utilizes numerical atomic
orbitals with Troullier-Martins pseudopotentials~\cite{Troullier91}.
We applied the high level double-$\zeta$ plus polarization functions
for both carbon and nitrogen atoms. In the {\sc SIESTA} calculations
no symmetry restriction was applied. The linear combination of atomic
orbitals (LCAO) analysis of the defect states is straightforward in
this methodology through the wavefunction coefficients obtained directly
from the {\sc SIESTA} calculations. In the {\sc VASP} calculations we
use a plane wave basis set with cutoff 420~eV($\approx$30~Ry) which is
adequate for well converged calculations using projected augmented
wave (PAW) pseudopotentials for the C and N atoms
\cite{Blochl94,Kresse99}. In the {\sc VASP} calculations we applied
the C$_{3v}$ symmetry, and the energy of the ground state as well as that of 
the excited states are calculated by setting the appropriate
occupation of the defect states in the gap. In the geometry
optimization calculations, all the atoms were allowed to relax until
the magnitude of the calculated forces was smaller than 0.01~eV/\AA . The
hyperfine tensor of the NV center was calculated by the {\sc CPPAW}
code~\cite{CPPAW}. In the {\sc CPPAW} calculations we used a 30~Ry
cutoff for the plane wave basis with PAW projectors which is virtually
equivalent with the methodology used in the {\sc VASP} calculations.

Convergence of calculated defect properties with supercell size is an important consideration.
For this reason, we have chosen to model the NV center by using a large 512-atom simple cubic
supercell. The lattice constant of the supercell ($\approx$14.2~\AA )
is four times larger than the lattice constant of the conventional
cubic cell of diamond, a$_0$=3.54~\AA. We used the $\Gamma$-point sampling in
the Brillouin-zone which corresponds to sampling finer than a $6\times
6\times 6$ grid of the primitive lattice; this provides a well
converged charge density. It is also advantageous to restrict the calculations to the
$\Gamma$-point in order to keep the degeneracy of the $e$ defect
states which may split in a general k-point sampling of the
Brillouin-zone. We checked that the geometry was practically identical
(to within 0.01~\AA ) going from a 216-atom fcc supercell to the
512-atom simple cubic supercell. We will show that the calculated spin
density, for instance, decays at much shorter distance than the
lattice constant of the supercell. Thus, the 512-atom supercell is adequate to 
represent the isolated NV defect in a realistic manner. 

We calculated the hyperfine tensor of the defect with the optimized
geometry obtained by the {\sc VASP} code. In the calculation of the
hyperfine tensor the relativistic effects are taken into account
\cite{Blochl00}. The hyperfine tensor of nucleus $I$ consists of the
Fermi-contact term (first parenthesis in the following equation) and
the dipole-dipole term (second parenthesis):
\begin{equation}
\label{eq:hyperf}
A^{(I)}_{ij} = \frac{1}{2S} \int {\text{d}^3\text{r} \text{ n}_\text{s}(\textbf{r})
\gamma_I \gamma_e \hbar ^2 \Big[\Big(\frac{8\pi}{3}\delta
(\text{r})\Big) + \Big(\frac{3 \text{x}_i \text{x}_j}{\text{r}^5} -
\frac{\delta _{ij}}{\text{r}^3}\Big)\Big] }
\end{equation}
where n$_\text{s}(\textbf{r})$ is the spin density of the spin state
$S$, $\gamma_I$ is the nuclear Bohr-magneton of nucleus $I$ and
$\gamma_e$ is the electron Bohr-magneton. The Fermi-contact term is
proportional to the spin density localized at the place of the nucleus
which is dominant compared to the dipole-dipole term. The ratio of the
Fermi-contact and dipole-dipole terms characterizes the shape of the
spin density. 

The contribution of $s$-like wave functions to the
charge density has a large effect on the Fermi-contact term but
negligible effect on the dipole-dipole term, since the $s$-like wave
function has a maximum at the positions of the nuclei and it is an
even function.  In contrast to this, the contribution of $p$-like wave functions to
the charge density has a negligible effect on the Fermi-contact term
but a large effect on the dipole-dipole term, since the $p$-like wave
function has a node at the place of nuclei and it is an odd function.
Typically, the contribution of the dipole-dipole term is significant
for the spin density built from well-localized dangling bonds, that
is, the \ $sp^3$ hybrid orbitals (see Table~\ref{tab:hyperf}). We note
that the pseudopotential methodology produces artificially smooth wave
functions close to the nuclei, therefore only the all-electron PAW
methodology can provide reliable hyperfine tensors. In the PAW
methodology the calculation of the hyperfine tensor is somewhat more
subtle than Eq.~\eqref{eq:hyperf} shows and the dipole-dipole term is
not fully calculated (see the appropriate note in
Ref.~\onlinecite{Blochl00}) which causes about 0.3~MHz inaccuracy in
the calculated dipole-dipole term in our case. This error could be
important to take into account for hyperfine tensors with small matrix
elements ($<$3~MHz). The total spin density, n$_\text{s}(\textbf{r})$,
can be defined as 
$$\text{n}_\text{s}(\textbf{r}) =
\text{n}_\text{up}(\textbf{r})-\text{n}_\text{down}(\textbf{r})$$ 
where $\text{n}_\text{up}(\textbf{r})$ and
$\text{n}_\text{down}(\textbf{r})$ are the spin densities built from
spin-up and spin-down electrons, respectively. Taking the
$M_\text{S}$=1 state of the $^3A_2$ state, as shown in
Fig.~\ref{fig:levels}, we expect that n$_\text{s}(\textbf{r})$ will be
positive. We will show that this is not true for the entire space
around the defect.

\section{Atomic and electronic structure of the NV defect}

\subsection{Geometry and electronic levels}

We begin with a discussion of the geometry optimization of the negatively charged NV
center obtained from the {\sc SIESTA} calculations using spin-polarization and
no symmetry restrictions. The defect automatically finds t the
S=1 state and maintains the C$_{3v}$ symmetry of the original, unrelaxed 
structure. The calculated
one-electron defect levels are shown in Fig.~\ref{fig:levels}. The
first neighbor C and N atoms clearly relaxed outward from the vacancy.
The calculated distances from the vacant site are 1.63 and 1.69~\AA\
for the C atoms and the N atom, respectively, that is,  the N atom relaxes 
more than the C atoms. We note that the C-vacancy distances are the
same within 0.0002~\AA\ without any symmetry constraints after
geometry optimization. The N-C bond lengths are 1.46~\AA, while the
bond lengths of C radicals are 1.50~\AA , which is not far from
1.44 and 1.45~\AA, respectively, obtained in an LDA molecular cluster
calculation~\cite{Goss96}. The localized basis sets can have problems
in the description of vacancy-like defects. Our {\sc VASP}
calculation, which employs a plane wave basis set, shows basically the
same geometry after optimization: the calculated distances from the
vacant site are 1.62 and 1.68~\AA\ for the C atoms and the N atom,
respectively. Thus, we conclude that the double-$\zeta$ plus polarization
basis provides results very close to those of the converged plane wave
basis set. 

In an earlier work by by \L uszczek \emph{et al.} \cite{Luszczek04},
a plane wave basis set with pseudopotentials was employed
in a 64-atom cubic supercell using $2\times 2\times 2$ Monkhorst-Pack
Brillouin Zone sampling~\cite{MP76} 
to investigate the NV defect in diamond. In that work, only the nearest
neighbor atoms to the vacant site were allowed to relax without symmetry restrictions and 
a geometry close to C$_{3v}$ symmetry was obtined; the largest deviation in
the C-vacancy distances was about 0.001~\AA . The calculated
distances from the vacancy were 1.67 and 1.66~\AA\ for the C atoms
and the N atom, respectively, which shows the opposite trend from what we
find both in the {\sc SIESTA} and in the {\sc VASP} calculations in the larger 
unit cell. Most probably the difference is due to the insufficient relaxation
condition restricted only to the first neighbor atoms around the
vacant site.

\begin{figure}[h]
  \centering
  \includegraphics[keepaspectratio,width=7.5cm]{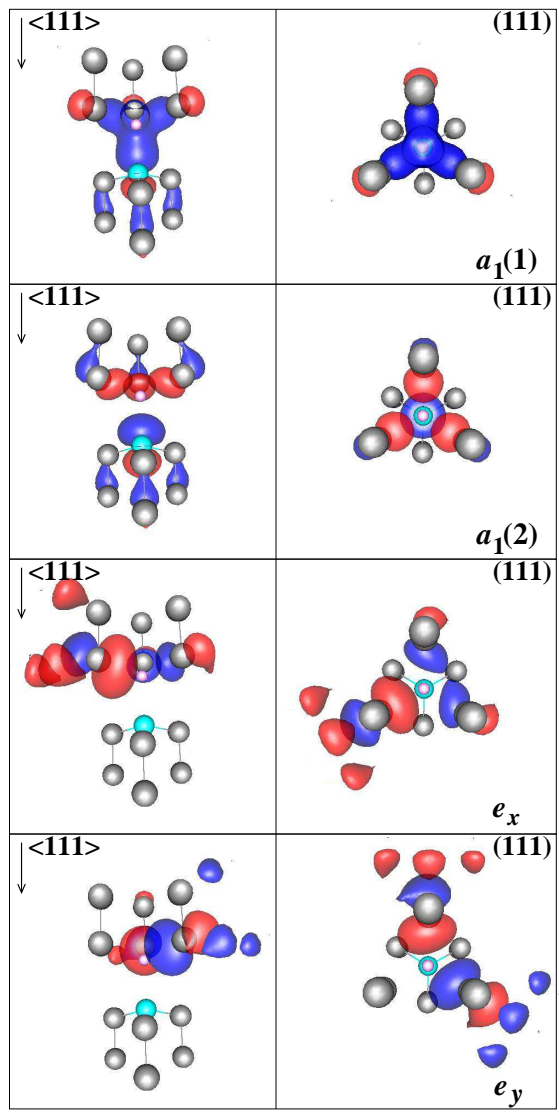}
  \caption{(Color online) Isosurfaces of the calculated wavefunctions
  of the $a_1(1), a_1(2), e_x, e_y$ defect states, shown in side (left)
  and top (right) views relative to the $<$111$>$ axis. Blue (red)
  isosurfaces correspond to negative (positive) values of the wavefunction. The
  small pink circle represents the position of the vacant site, while the
  grey and cyan balls show the C and N atoms,
  respectively. We show the atoms up to the second neighbor from
  the vacant site.}
  \label{fig:wfs}
  \centering 
\end{figure}

\begin{figure}[h]
  \centering
  \includegraphics[keepaspectratio,width=15cm]{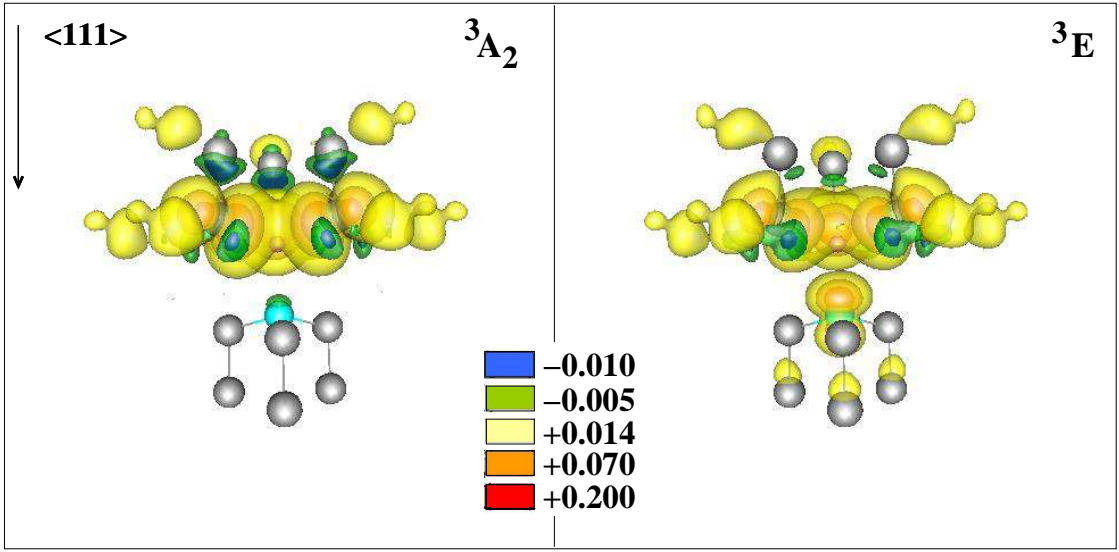}
  \caption{(Color online) Calculated spin density isosurfaces in the
    $M_\text{S}$=1 state for the $^3A_2$ (left) and the $^3E$ state (right).  
    The vacant site is depicted by a small pink sphere at the center 
    of each plot.}
  \label{fig:spin3E}
  \centering 
\end{figure}

We plot the wave functions of the defect states obtained by the LSDA
calculations in Fig.~\ref{fig:wfs}.  
The group theory analysis based on the single-particle picture
describes very well the defect states. Naturally, the defect states
are not strictly localized on the first neighbor atoms of the vacancy
but the largest portion of the wave functions can be indeed found
there.  The {\sc SIESTA} calculation gives $\alpha \approx 0.7$ for the 
parameter that appears in 
Eq.~\eqref{eq:orbitals}. This means that the N orbital is
mostly localized on the $a_1(2): \phi _2$ defect level and has no
amplitude on the $e$ levels. Therefore, the nitrogen atom is only very
weakly spin-polarized in the $^3A_2$ state, while it is strongly
spin-polarized in the $^3E$ state (when one electron is excited from
the $a_1(2)$ to level to the $e$ level). This is clearly shown on
Fig.~\ref{fig:spin3E}. It is apparent, from this figure, that the N atom is only
weakly polarized (small negative spin density) in the $^3A_2$
state while it is strongly polarized in the $^3E$ state comparable to
the C ligands (large positive spin density). The spin density is always highly
localized on the three C atoms around the vacant site (orange
lobes in the figure).

We also checked the situation when we optimized the geometry with the
condition S=0. We already showed that the
$|\sigma_2\bar{\sigma_2}\sigma_3\bar{\sigma_3}\rangle$ state is not an
eigenstate with C$_{3v}$ symmetry. In addition, this state is a
Jahn-Teller unstable system. Indeed, the defect reconstructs to
C$_{1h}$ symmetry to remove the degenerate $e$ level. However, this
configuration is about 0.3~eV higher in energy than the $^3A_2$ state
with C$_{3v}$ symmetry.

\subsection{Energy sequence of multiplets}

\begin{figure}[h]
  \centering
  \includegraphics[keepaspectratio,width=8cm]{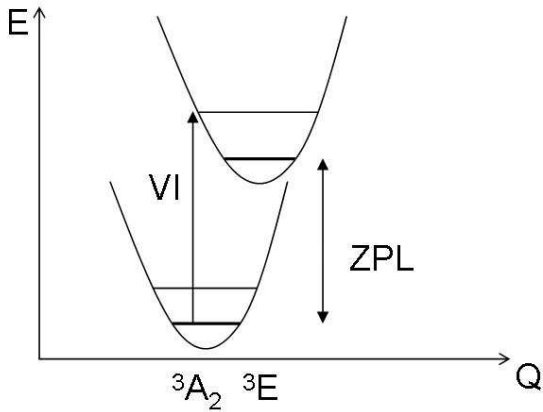}
  \caption{Energetics of photoluminescence absorption: VI is the vertical
  ionization energy, ZPL is the zero-phonon line transition, E is the total
  energy and Q the configuration coordinate. }
  \label{fig:PL}
  \centering 
\end{figure}

From the structural analysis, we conclude that the dangling bonds
around the vacancy do not form long bonds which could be the driving
force of the reconstruction. Instead, the atoms relax outward from the
vacancy and retain the strongly localized dangling bonds pointing to
the vacant site which maintains the C$_{3v}$ symmetry. Since the
degenerate $e$ defect level is only partially occupied, this is a
typical situation where configurational interaction plays a crucial
role. As shown above, most of the singlet eigenstates can be described
only by multi-determinantal wavefunctions in C$_{3v}$ symmetry. The
optical transition takes place between the triplet states. We already
showed the results on the $^3A_2$ state. In the {\sc VASP} calculation
it is possible to set the occupation of one-electron states. The $^3E$
state can be achieved by setting zero occupation for the spin-down
$\phi _2$ level and full occupation of $\phi _3$ spin-up and spin-down
levels. The energy of the $^3E$ state can be calculated in the fixed
geometry of the $^3A_2$ state which yields the vertical ionization
energy. Upon the excitation of the electron the nuclei can relax to
find the minimum energy in the new configuration space. This
relaxation can take place with the help of phonons around the defect.
The ZPL transition corresponds to that energy where phonons do not
participate, between the energy minima of the two configurations as
shown in Fig.~\ref{fig:PL}.

The calculated vertical ionization energy is 1.91~eV within
LSDA. We found that the NV defect significantly relaxes due to this
internal ionization. The C-vacancy distance is 1.67~\AA\ while the
N-vacancy distance is 1.61~\AA\ in the $^3E$ state. This shows the
opposite trend than what was found in the $^3A_2$ state. This may be
understood as follows: the N atom is strongly spin-polarized in the
$^3E$ state compared to $^3A_2$ state, while the C ligands will be
somewhat less spin-polarized which induces different charge transfer
between the atoms in the $^3E$ state, and leads to a different geometry.
The calculated relaxation energy (the Fracnck-Condon shift) is 0.2~eV. 
From this, we find a ZPL energy of 1.71~eV which
can be tentatively compared to the experimental value of 1.945~eV
\cite{duPreez65,Davies76}. We note that a similar value (1.77~eV) was
found by the LSDA molecular cluster calculation~\cite{Goss96}. The
LSDA excitation energy and the experimental transition energy are
remarkably close to each other. This shows that the self-interaction
error of LSDA for these defect levels does not differ too much, which
is not unexpected since both of the defect states are basically
valence band derived (from $sp^3$-like hybrid orbitals). Nevertheless,
the calculated Franck-Condon shift, which is the relaxation
energy defined as the energy difference between
the vertical ionization energy and the ZPL energy, should be even more accurate than
the calculated internal ionization energy. Indeed, the PL spectrum
shows a broad phonon spectrum even at low temperature, and the
intensity of the ZPL line is relatively small compared to the phonon
side bands which indicates a large Franck-Condon shift.

In addition to the triplet states it is worthwhile to calculate the
energies of the singlet states because they play a significant role in
the emission process but these states have not yet been measured
directly in experiments (see Ref.~\onlinecite{Manson06} and references
therein). LSDA is not a suitable methodology to calculate these
energies accurately. Beside the self-interaction error (which is
relatively small for these defect levels as discussed above), LSDA
gives the charge density of the interacting electrons, which
is expressed in terms of the non-interacting Kohn-Sham particles. In
the Hartree-Fock language the total wave function, which is not
directly used in DFT-LSDA, is expressed by a single Slater-determinant.
Thus, the usual DFT-LSDA calculation cannot represent most of the
singlet eigenstates of the NV center.  This can be corrected by using
a perturbation theory within DFT-LSDA as explained by Lannoo \emph{et
  al.}~\cite{Lannoo81}. Instead, we adopt here a more approximate
method~\cite{Barth79} which was already applied for some states of
this defect in Ref.~\onlinecite{Goss96}. Our goal is to give
approximate energies of the singlet eigenstates in order to predict
their {\em sequence}, and we do not attempt to provide energies directly
comparable to the experiments. 

One of the $^1A_1$ states can be described by a single
Slater-determinant (when the $a_1(2)$ level is totally empty and the
$e$ level is fully occupied). This state can be calculated with
geometry optimization by LSDA as explained for the $^3E$ state.
We therefore concentrate on the remaining $^1A_1$ and two $^1E$ states. While
the $|\sigma_2\bar{\sigma_2}\sigma_3\bar{\sigma_3}\rangle$,
$|\sigma_2\bar{\sigma_2}\sigma_3\bar{\sigma_4}\rangle$, and
$|\sigma_2\sigma_3\bar{\sigma_3}\bar{\sigma_4}\rangle$ single
Slater-determinants are not eigenstates, they can be expressed as the
linear combination of different eigenstates shown in
Eqs.~\eqref{eq:multi1a},~\eqref{eq:multi1b} and \eqref{eq:multi2}. von
Barth~\cite{Barth79} has shown that the LSDA total energy ($\Et$) of the mixed state
can be expressed as the appropriate sum of the energy of the
eigenstates, which yields the following equations:
\begin{subequations}
\begin{equation}
\label{eq:energya}
\Et[|\sigma_2\bar{\sigma_2}\sigma_3\bar{\sigma_4}\rangle] =
 \frac{1}{2}(\Et[^1E] + \Et[^3A_2])
\end{equation}
\begin{equation}
\label{eq:energyb}
\Et[|\sigma_2\bar{\sigma_2}\sigma_3\bar{\sigma_3}\rangle] = \frac{1}{2}(\Et[^1E] + \Et[^1A_1])
\end{equation} 
\begin{equation}
\label{eq:energyc}
\Et[|\bar{\sigma_2}\sigma_3\sigma_4\bar{\sigma_4}\rangle] = \frac{1}{2}(\Et[^3E] + \Et[^1E])
\end{equation}
\end{subequations}      
In Eqs.~\eqref{eq:energya} and \eqref{eq:energyc} we assume that the
energy of the triplet states with $M_\text{S}=0$ and $M_\text{S}=1$ is
the same. This is a very good approximation since, for instance, the
experimentally measured splitting is about 2.88~GHz (few $\mu$eV) for
the ground state due to spin-spin interaction \cite{Loubser77}, while
the spin-orbit splitting for the $^3E$ state is expected to be within
few meV, which is far beyond the accuracy of LSDA calculations. The
energy of the mixed states on the left hand side of
Eqs.~\eqref{eq:energya}-\eqref{eq:energyc} can be calculated directly
by LSDA. Since $\Et[^3A_2]$ is known, $\Et[^1E]$ can be determined
from Eq.~\eqref{eq:energya}. Similarly, $\Et[^1E]$ can be determined
from Eq.~\eqref{eq:energyc}. By combining Eqs.~\eqref{eq:energya} and
\eqref{eq:energyb} we arrive at:
\begin{equation}
\label{eq:energya1}
 \Et[^1A_1] = 2
 (\Et[|\sigma_2\bar{\sigma_2}\sigma_3\bar{\sigma_3}\rangle] -
 \Et[|\sigma_2\bar{\sigma_2}\sigma_3\bar{\sigma_4}\rangle]) +
 \Et[^3A_2]) 
\end{equation}
Thus, the $^1A_1$ state can be also determined. 

von~Barth~\cite{Barth79} applied this approach successfully to calculate the energy
of atoms in different states. Formally, this method can
also be applied to the NV defect in diamond, but attention must be
paid to relaxation effects. Generally, if the electron state changes
then it may imply also relaxation of the ionic positions, as was the
case for the $^3E$ state discussed earlier.  Relaxation effects cannot
be taken into account with this methodology, since only the energy of
the mixed states can be calculated directly by LSDA, and the
relaxation of the mixed state is meaningless.  In other words, the
geometry must be fixed in these calculations. The occupation of the
$e$ state varies in the case of the $^1A_1$ and $^1E$ states of the
$a_1^2(2)e^2$ configuration. We assume that the geometry would involve
negligible change from the geometry of the $^3A_2$ state which belongs
also to the $a_1^2(2)e^2$ configuration. Therefore, we fix the
geometry obtained in the $^3A_2$ state in the process of calculating the
$^1A_1$ and $^1E$ states of the $a_1^2(2)e^2$ configuration. Using the
same argument we fix the geometry obtained for the $^3E$ state in the
calculation of the $^1E$ state of the $a_1^1(2)e^3$ configuration.
With these, we find the following energy sequence of the multiplets:
$$\Et[^3A_2] \xrightarrow{\approx 0.0~eV}
\Et[^1A_1] \xrightarrow{\approx 0.9~eV} \Et[^1E] \xrightarrow{\approx
  0.8~eV} \Et[^3E] \xrightarrow{\approx 0.5~eV} \Et[^1E]
\xrightarrow{\approx 1.3~eV} \Et[^1A_1],$$ 
that is, the deeper $^1A_1$ state is close in energy to the $^3A_2$ state,
and the deeper $^1E$ state is below $^3E$ state.  The energy
differences between the triplet states were already discussed.

We focus next on the singlet states. Within our approximate methodology
the $^3A_2$ and $^1A_1$ states are almost degenerate.  Close
inspection of Eq.~\eqref{eq:energya1} reveals that the energy sequence
of the $^3A_2$ and $^1A_1$ states depends on the energy difference of two
singlet states:
$(\Et[|\sigma_2\bar{\sigma_2}\sigma_3\bar{\sigma_3}\rangle] -
\Et[|\sigma_2\bar{\sigma_2}\sigma_3\bar{\sigma_4}\rangle])$. We obtain
almost zero for this energy difference. However, the LSDA
self-interaction error may be larger for the
$|\sigma_2\bar{\sigma_2}\sigma_3\bar{\sigma_3}\rangle$ state (where
the exchange energy of $\sigma_3$ appears) than for the
$|\sigma_2\bar{\sigma_2}\sigma_3\bar{\sigma_4}\rangle$ state (where
the electrons occupy spatially orthogonal orbitals). This may raise
the energy of the $\Et[^1A_1]$ state. Nevertheless, this can be
partially compensated by the relaxation effect of the $^1A_1$ state
which we are neglecting by necessity as explained before. 
An additional issue is the possible mixing with the
higher $^1A_1$ state.  We argue that the two $^1A_1$ states are not
likely to mix because they are very far from each other in energy.

We conclude from the above analysis 
that the $^1A_1$ state is indeed close in energy to the $^3A_2$ state.
This may imply a very complicated fine structure of the
states. Taking the spin-spin interaction into account the $^3A_2$
state splits to $A_1(M_\text{S}=0)$ and $E(M_\text{S}=\pm 1)$ states,
while $^1A_1$ becomes $A_1(M_\text{S}=0)$ (the spin state is not a
good quantum number anymore, just its projection). If the two
$A_1(M_\text{S}=0)$ states are close in energy then they
may mix with each other. It is also important to notice that the
energy of the deeper $^1E$ state falls between the energies of the triplet
states. The relaxation effect may lower the energy of this state. In
addition the two $^1E$ states are not far from each other in energy,
so the off-diagonal elements in the hamiltonian may not be neglected.
The mixing of the two $^1E$ states would further lower the energy of
the deeper $^1E$ state and would raise the energy of the higher $^1E$
state. The final conclusion is that there are two singlet states
between the triplet states, and the $^1A_1$ state is much closer in
energy to the ground state than the $^1E$ state. The energies of the
other two singlet states are certainly above that of the $^3E$ state.

The deeper singlet $^1A_1$ and $^1E$ states can play an important role
in the emission process of the NV center.  Experiments indicate that
there should be a possibly long-living singlet state between
the triplet states (see Ref.~\onlinecite{Manson06} and references
therein). Usually, the singlet $^1A_1$ is considered in this process.
However, our calculations indicate that there are two singlet states
between the triplet states.  Goss \emph{et al.}~\cite{Goss96} reported the sequence
of $^3A_2$, $^1E$, $^1A_1$, and $^3E$ states which is surprising in
light of the previous discussion, as these states were obtained by
LSDA molecular cluster calculations. Manson \emph{et
  al.}~\cite{Manson06} recently showed that if the $^1E$ state is above
the $^1A_1$ state, the known properties of the emission can be
consistently explained as with the original singlet $^1A_1$ model,
with the only difference being that the $^1E$ and $^1A_1$ states both
contribute to the spin polarization process during the optical cycling
and that the effect will be more efficient.  Having the $^1A_1$ state
higher in energy than the $^1E$ state would result in no change in
spin orientation during optical cycling which is in contradiction with
experiment~\cite{Manson06}. From this point of view our results
obtained from approximate calculations are consistent with the PL
experiments. 

\section{Hyperfine constants}

As mentioned in the Introduction, the NV center in diamond
is a promising candidate to realize qubit solid state devices
operating at room temperature (Ref.~\onlinecite{Childress07} and
references therein). The qubit is the non-zero (S=1) electron spin
ground state which can interact with the neighbor $^{13}$C isotopes
possessing $I$=1/2 nuclear spin via hyperfine interaction. The natural
abundance of the $^{13}$C isotope is about 1.1\%, so we can assume the
same abundance in the diamond lattice. In conventional EPR
measurements, the EPR absorption signal is detected on the
\emph{ensemble} of the defects in the diamond sample. The sample
should be thick enough for absorption measurements and the
concentration of the defects should be sufficiently high. Finally, a
large number of defects is measured at the same time by EPR, so
statistics can be applied to analyze the data. If the spin density is
strongly localized on three \emph{symmetrically equivalent} C ligands
(see Fig.~\ref{fig:spin3E}) then the probability of finding one
$^{13}$C atom among them is given by the binomial distribution
and is about 3.2\%. Finding two or three of them has
negligible probability.  Due to the $I$=1/2 nuclear spin, the hyperfine
interaction splits to two lines with $I_z = 1/2$ and $I_z = -1/2$ and 
therefore the intensity ratio between the
main hyperfine line (involving no hyperfine interaction with $^{13}$C
atoms) and the hyperfine line associated with the C ligands will be
roughly 1.5\%. 
This makes the EPR measurement on $^{13}$C hyperfine interaction a
challenging task, since the signal to noise ratio should be very good
and the intensity of the EPR signal should be strong enough and stable
to identify the satellite hyperfine lines due to $^{13}$C isotopes.

The $^{13}$C hyperfine interaction has been detected in EPR from two
sets of three symmetrically equivalent C atoms by Loubser and van Wyk
\cite{Loubser77,Loubser78}. The larger hyperfine constants were
associated with the C-ligands of the NV center.  In addition, the
hyperfine interaction of the $^{14}N$ isotope was found in NV center
by EPR and ENDOR measurements~\cite{Loubser77,He93}.  To our
knowledge, hyperfine interaction with other $^{13}$C isotopes has not
been measured directly by EPR.  We can estimate the localization of
the charge density on the C and N atoms that are immediate neighbors
of the vacant site, from the linear combination of the atomic orbitals as
they appear in the wavefunctions.  He {\it et al.}~\cite{He93}
estimated that 72\% and 0.2\% of the charge density is localized on the
three C ligands and the N atom, which leaves approximately 28\% of the
charge density to be spread in the lattice. Wrachtrup {\it et al.}
\cite{Wrachtrup01} speculated that the spin density decays
exponentially as a function of the distance from the vacant site.
Since the hyperfine constants are roughly proportional to the spin
density, as Eq.~(\eqref{eq:hyperf}) indicates, these authors proposed
that 9 or more carbon nuclei should have a hyperfine value of 70~MHz
in the second neighborhood while the more distant carbon atoms should
have a hyperfine constant smaller than 10~MHz.  This proposal is not
entirely consistent with the known EPR data, since isotropic hyperfine
splitting of 5.4~G($\approx$15~MHz) was measured from three $^{13}$C
isotopes~\cite{Loubser78} whereas a value of 70~MHz hyperfine
splitting should be measurable by EPR because it would not be
obscured by the main EPR line. 

In their theoretical treatment of the NV center, \L uszczek {\it et al.}
\cite{Luszczek04} claimed that they can support the proposal of Wrachtrup {\it et al.}
\cite{Wrachtrup01} based on their \emph{ab initio}
results. They optimized the geometry without
symmetry constraints only for the first neighbor atoms of the vacancy
in a 64-atom supercell. The adequacy of this restriction was already
discussed above; with this restriction, while the C$_{3v}$ symmetry is
almost retained, the calculated Fermi-contact hyperfine interactions
for the three C ligands deviate from each other more than 10\% (see
Table 2 in Ref.~\onlinecite{Luszczek04}). This suggests that the spin
density was not adequately converged in that calculation since the small
deviation in the geometry from the C$_{3v}$ symmetry could not imply
such a large discrepancy in the calculated hyperfine field. These
authors also calculated the Fermi-contact hyperfine interaction for
the C atoms situated about 2.5~\AA\ away from the vacant site. The
reported numbers were about an order of magnitude smaller than for the
C ligands, which led to their conclusion that the spin density and the
corresponding hyperfine constants should decay fast for other C-atoms
farther from the vacant site.  

In addition to the problem of the
inconsistent values of the hyperfine interaction for the three
C-ligands, several other issues related to the results of
Ref.~\onlinecite{Luszczek04} must be mentioned:\\
(i) These authors reported actually the hyperfine \emph{field} and not
the hyperfine \emph{constant}.  However, the conversion from the hyperfine
field, which is the magnetization density on the atom and it is a
number directly obtained from the computation, to the hyperfine constant is not
unique; therefore, it is very difficult to compare the calculated values to the
experimental data.\\
(ii) Only the Fermi-contact term was calculated while the
dipole-dipole term can be also significant; this is known to be the
case for the C ligands from experimental measurements
\cite{Loubser77,Loubser78}.\\
(iii) The hyperfine interaction with distant C atoms could be
very important for qubit applications based on this defect, so the
hyperfine tensor must be calculated at larger distances from the
vacancy.\\
Based on these arguments, we believe that the nature of the spin density and the
corresponding hyperfine interaction with the $^{13}$C isotopes has not
yet been explored in detail despite of its high importance.

\begin{figure}[h]
  \centering
  \includegraphics[keepaspectratio,width=15cm]{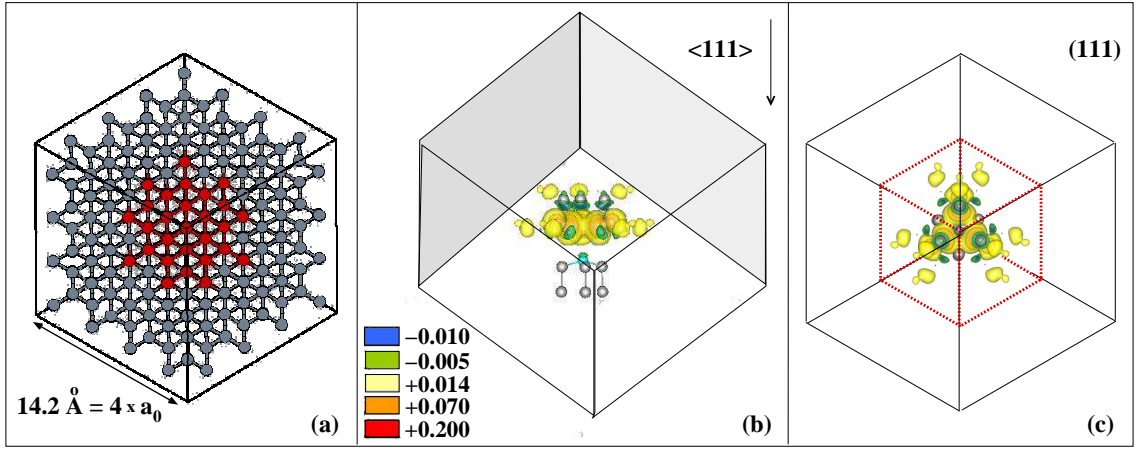}
  \caption{(Color online) 
    (a) The 512-atom cubic supercell with all C atoms shown (atoms within a 64-atom
    supercell are shown in red).  The suprercell size in \AA~ and in units of the conventional 
    cubic cell lattice constant, $a_0$, is indicated on the side of the cube.     
    (b) Perspective view of the calculated spin density isosurfaces in the
    $M_\text{S}$=1 state. Only atoms up to
    the second nearest neighbor of the vacant site are shown.
    (c) View along the (111) direction indicating the C$_{3v}$ symmetry 
    of the spin density, which is given in colored contours. 
    The black lines denote the size 
    of the 512-atom supercell, while the red dotted ones show the 
    boundaries of the 64-atom supercell.
}
  \label{fig:spin3A2}
\end{figure}

Toward establishing the nature of this interaction, we show first the
calculated spin density in our 512-atom supercell in
Fig.~\ref{fig:spin3A2}.
As expected, the spin density is highly localized around the three
C-ligands nearest to the vacant site (orange lobes in the figure),
within a radius of $1\times \text{a}_\text{0}$ from this site.
The spin density practically vanishes at distances $>2\times \text{a}_\text{0}$
from the vacant site. Below the three C atoms (see Fig. \ref{fig:spin3A2})
there is the N atom with a small \emph{negative}
    spin density. Some C atoms farther from the vacant site also
    have negative spin density. The spin density extends mostly on a
    plane perpendicular to the (111) direction and no measurable spin
    density can be found below the N atom. On
the N atom, the spin density is negative. It was shown earlier that
the spin density comes mostly from the spin-polarized $e$ level
localized on the C dangling bonds, and due to symmetry reasons,
orbitals related to the N do not appear in the $e$ level.
The tiny negative charge density on the N atom can be explained
by the polarization of its core states: since the nuclear
Bohr-magneton of the N atom is positive, the Fermi-contact term
will be negative (see Table~\ref{tab:hyperf}). Overall, the spin
density is spread on a plane perpendicular to the (111) direction.
There are some C atoms which have significant negative spin
density which results in negative Fermi-contact term.  No measurable
hyperfine interaction (spin density) can be found for the C atoms
below the N atom.  Loubser and van Wyk~\cite{Loubser78}
speculated that the 15~MHz $^{13}$C isotropic hyperfine splitting
comes from the three C atoms bonded to the N atom. Our
calculation negates this possibility.

\begin{table}
\caption{\label{tab:hyperf} The calculated principal values of the
  hyperfine tensor (columns 3 to 5) compared to the known experimental
  data (columns 6 to 8) in MHz. The average of the three principal
  values yields the Fermi-contact term. The difference between the
  principal values and the Fermi-contact term gives the dipole-dipole
  term.  Only atoms with a signal larger than 2~MHz are shown.
  The symmetrically equivalent number of C atoms is shown in
  the first column and their distance in \AA~ from the vacant site in the second
  column. The experimental data on $^{14}$N is taken from
  Refs.~\onlinecite{Loubser78,He93}. Experimental data on $^{13}$C
  atoms were taken from Ref.~\onlinecite{Loubser78}. EPR
  studies can directly measure only the absolute value of the
  hyperfine constants, which is indicated by adding a $\pm$ sign to 
  experimental values. The
  calculated hyperfine tensors can be used for comparison with
  spin-echo measurements (see text).}
\begin{ruledtabular}
\begin{tabular}{cccccccc}

Atom 
  & R$_\text{vac}$ & A$_{11}$ & A$_{22}$ & A$_{33}$ & A$^{exp}_{11}$ &
  A$^{exp}_{22}$ & A$^{exp}_{33}$\\ \hline
$^{14}$N &  1.68  & -1.7  &   -1.7  &   -1.7  &  $\pm$2.1 & $\pm$2.1 & $\pm$2.3 \\  
 3 C   &    1.61  &  109.5&    110.2&   185.4 &  $\pm$123 & $\pm$123 & $\pm$205 \\
 6 C   &    2.47  &   -4.8&     -3.7&    -1.5   &           &          &          \\ 
 3 C   &    2.49  &   -7.4&  -7.3& -5.8 &  & & \\  
 6 C   &    2.90  &    2.8&      3.3&     4.6  &  & & \\
 3 C   &    2.92  &    1.4&      2.4&     2.9 &  & & \\
 3 C   &    2.93  &    3.4&      4.7&     4.9 &  & & \\
 6 C   &    3.85  &   13.5&    14.2 &    19.4 &  & & \\
 3 C   &    3.86  &   12.8&    12.8 &    18.0  &  $\pm$15.0 & $\pm$15.0 & $\pm$15.0\\
 6 C   &    4.99  &    2.6&      2.7&     3.8 & & & \\
 3 C   &    5.00  &    1.5&      1.5&     2.2 & & & \\ 
\end{tabular}
\end{ruledtabular}
\end{table}
Table~\ref{tab:hyperf} shows that there is a local maximum of the spin
density at R$_\text{vac}\approx$3.9 and 5.0~\AA , where R$_\text{vac}$
is the distance from the vacant site. At R$_\text{vac}>$6.3~\AA\ the
calculated hyperfine constants are below 1~MHz which means that the
spin density vanishes at R$_\text{vac}\approx 2\times$a$_\text{0}$.
Apparently, the 64-atom supercell is too small to capture these properties
due to the artificial overlap of the spin density caused by the
periodic boundary conditions.

\begin{figure}[h]
\centering
\includegraphics[keepaspectratio,width=8cm]{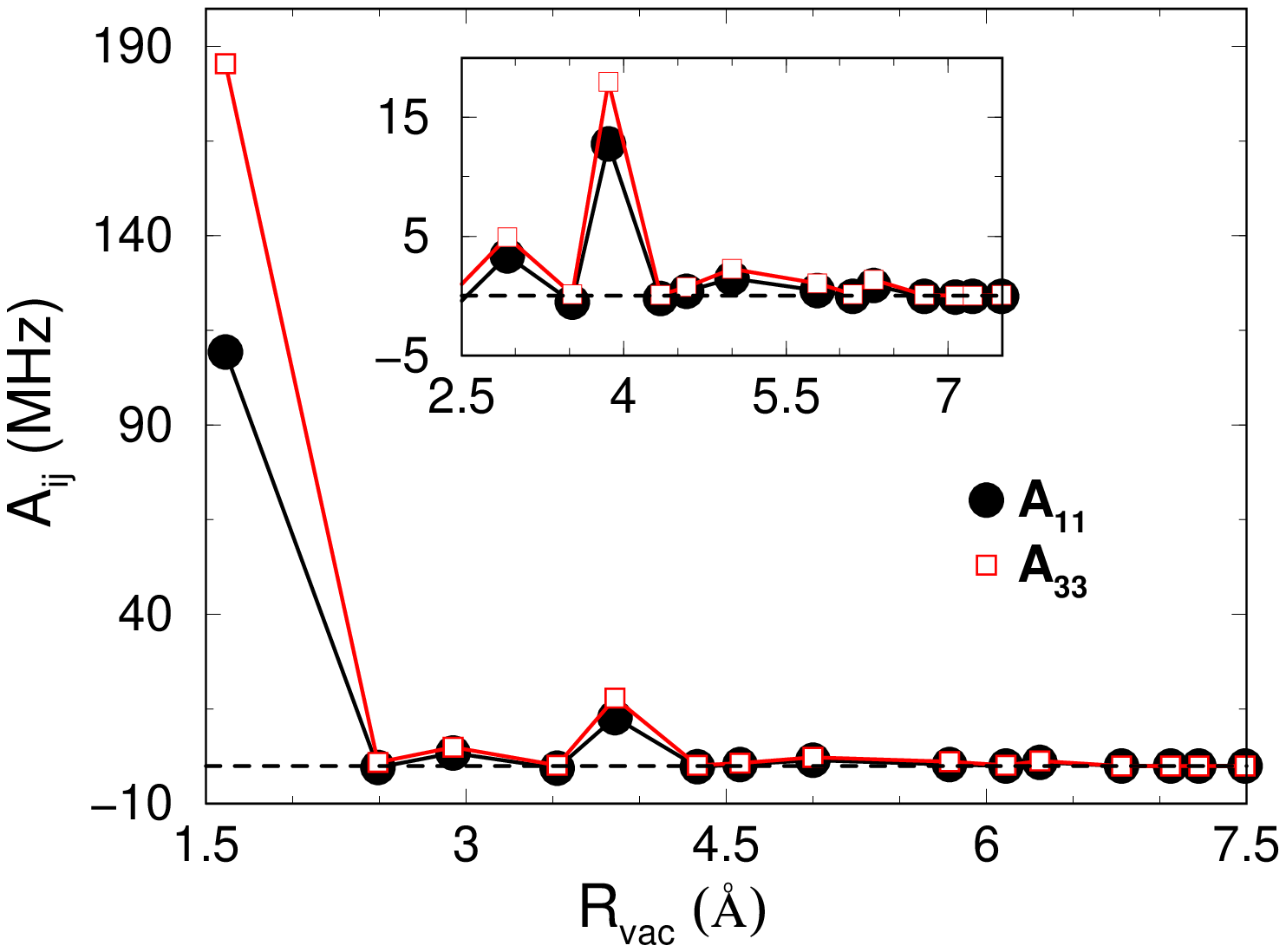}
\caption{Variation of the principal values of the hyperfine tensor
  A$_{11}$ and A$_{33}$ as a function of the distance from the vacant site
  ($R_{vac}$) for the set of 3 symmetrically equivalent C atoms (3C).
  In the inset we show changes farther from the vacant site on a finer scale.
  The variation for the set of 6 symmetrically equivalent C atoms
  follows closely the one shown for 3C.}
\label{fig:hyperf}
\centering  
\end{figure}

We also find that the charge density does not decay monotonically like an
exponential function, but behaves more like a wavepacket, that is,
as $f(x) = \sin(x)/x$ (see Fig.~\ref{fig:hyperf}).  

It is important to compare the calculated hyperfine values with the
known experimental data in order to estimate the accuracy of our
calculations. For that purpose, the hyperfine constant of the N
atom is not the best choice because its value is very small and it is
caused by only indirect spin-polarization effect. As mentioned above,
there is an inherent inaccuracy in the calculated dipole-dipole term
of about 0.3~MHz, therefore it is reasonable to consider only values of the
hyperfine constants that are significantly larger than this limit.
Accordingly, we restrict the comparison to values that are higher than
2~MHz.  By comparing the hyperfine constants of the C-ligands we
estimate the inaccuracy for both the Fermi-contact term and the
dipole-dipole term to be about 10\%. This is usual for LSDA
calculations~\cite{UmedaPRB05,SonPRL06,UmedaPRL06,UmedaPRB07}.  The
calculated ratio between the Fermi-contact term and the dipole-dipole term
agrees remarkably well with experiment for the C-ligands. This shows
that the shape of the spin density is very well described by LSDA.
This ratio indicates that the $p$-functions dominate by about 90\% in
the dangling bonds, so those are more $p$-like orbitals than $sp^3$
hybrids. The reason is most likely the outward relaxation of the
C-atoms from the vacant site. The plotted wavefunctions indeed show a
very strong $p$-contribution of the $e$ levels in Fig.~\ref{fig:wfs}
while the spherical $s$-contribution is very small. 
The shape of the wave functions can be directly compared
to the proposed wave functions in Eq.~\eqref{eq:orbitals} from the
group theory analysis of the defect diagram.

An additional 15~MHz $^{13}$C hyperfine splitting was measured by Loubser
{\it et al.}~ \cite{Loubser78} using EPR but the measured
spectrum of the NV center was not shown in detail, therefore we
cannot comment on the accuracy of this measurement. Nevertheless, that
work stated that the hyperfine splitting is isotropic, that is, the
dipole-dipole term is negligible, and the relative intensity of the
hyperfine satellite line and the main EPR line indicates the
involvement of three symmetrically equivalent C atoms
\cite{Loubser78}. The likely candidate for this signal is found at the
third neighbors of the vacant site at R$_\text{vac}$=3.86~\AA . The
spin density of these atoms is shown as yellow lobes above the small
green lobes at the highest position on the side view in
Fig.~\ref{fig:spin3A2}. The calculated anisotropy of this hyperfine
interaction is about 3-4~MHz. If the uncertainty in the measurement is
in this range due to line broadening, then the signal could be detected
as nearly isotropic. However, our calculations reveal that six
additional C atoms have similar hyperfine splitting at the third
neighbor distance of R$_\text{vac}$=3.85~\AA, corresponding to the six
yellow lobes laterally spread farthest from the vacant site, which is
most obvious from the view along the (111) direction in
Fig.~\ref{fig:spin3A2}. This means that the hyperfine splitting due to
these six C atoms would have to be detected simultaneously with the
other three ones. Since the difference in the hyperfine splitting of
these two sets of atoms is small, the six-atom set could obscure the
signal of the three-atom set showing an effective relative intensity
associated with six symmetrically equivalent C atoms.  We suggest that
this part of the spectrum should be re-investigated in detail
experimentally.  The hyperfine splitting of 7-8~MHz from the $^{13}$C
atom may be also detectable by EPR while the other signals may be too
small and hence obscured by the main hyperfine lines.

%
%
Recently, individual NV centers have been detected by spin-echo
measurements~\cite{Childress06,Childress07}.  In particular, detailed
results for six NV centers in diamond have been reported
\cite{Childress06}.  The spin-echo measurements have detected the
coherent state of the electron spin coupled with a proximal $^{13}$C
nucleus. The coupling is due to hyperfine interaction between the
electron spin and the nuclear spin of $^{13}$C isotopes. The resulting
spin-echo signals show a rapidly oscillating function enveloped by a
more slowly oscillating function~\cite{Childress06}.  These authors
proposed a theory to explain this signal, and they concluded that the
fast modulation frequency is due to the effective magnetization
density of the electron spin felt by the $^{13}$C nucleus, which is 
the same as the hyperfine interaction. The modulation frequency can
be well approximated as the norm of the hyperfine tensor projected to
the symmetry axis, which leads to the following expression within our
formulation of the problem:
\begin{equation}     
\label{eq:modfreq}
\omega ^{(X)} = \sqrt{\big( [1 1 1] \times A^{(X)}_{ij} \big)^2} 
\end{equation}
where $[1 1 1]$ is the appropriate projection vector and
$A^{(X)}_{ij}$ is the hyperfine tensor of nucleus $X$. Since we
calculate the full $A^{(X)}_{ij}$ tensor (see Eq.~\eqref{eq:hyperf}),
the calculation of the modulation frequency is straightforward.
Modulation frequencies have been reported for four single NV defects
(see Fig.~4B in Ref.~\onlinecite{Childress06}). A single modulation
frequency was measured for each NV center at $\approx$2~MHz,
$\approx$4~MHz, $\approx$9~MHz, and $\approx$14~MHz, respectively, so
these particular NV centers had hyperfine interaction with one
$^{13}$C isotope in the lattice~\cite{Childress06}. It would be useful
to compare the hyperfine interaction measured by EPR and spin-echo
techniques which can be an additional validation of the theory
developed by Childress and co-workers~\cite{Childress06}. As was
explained earlier, conventional EPR tools have limitations in
detecting $^{13}$C isotopes. A $^{13}$C enriched sample would be
useful to study this defect experimentally in more detail. 

The largest $^{13}$C hyperfine splitting corresponds to a modulation
frequency that is too large to be detected by spin-echo measurements.
However, an isotropic $^{13}$C hyperfine splitting of 15~MHz has also
been reported~\cite{Loubser78}. The isotropic signal means that the
modulation frequency should be also about 15~MHz. This is very close
to one of the measured modulation frequencies at $\approx$14~MHz
\cite{Childress06}. From the calculated hyperfine tensors in
Table~\ref{tab:hyperf}, this signal must originate from a C-atom which
is a third neighbor of the vacant site at R$_\text{vac}$=3.86~\AA .
The calculated modulation frequency is $\approx$16~MHz which is close
to the measured one taking into account the limitations of our computational method.
The 9~MHz spin-echo signal can originate only from the
atoms at R$_\text{vac}$=2.49~\AA , and contributions from other
neighbors can be safely excluded.  In this way, the origin of the
signal can be identified. The 4~MHz spin-echo signal can originate
either from atoms at R$_\text{vac}$=2.90~\AA\ or at
R$_\text{vac}$=2.93~\AA\, taking into account computational limitations. In the
first case, six symmetrically equivalent C atoms are involved, while
in the second case a set of three symmetrically equivalent C atoms are
involved. It is difficult, if not impossible, to identify the origin
of the 1~MHz spin-echo signal, which is beyond the accuracy of our calculations.
Nevertheless, the calculations indicate that this signal could arise
from at least 12 C atoms. Most of them are far from the vacant site
but some are closer than the atoms that give rise to the
$\approx$14~MHz signal, as is evident from Figure~\ref{fig:hyperf}.

The number of
symmetrically equivalent atoms is also specific to the
individual hyperfine constant, and therefore to the modulation
frequency. Because of the C$_{3v}$ symmetry, sets of three or sets of six C atoms are
equivalent with each other. The relative occurrence of the modulation
frequencies measured by spin-echo experiments helps in identifying
the equivalent atoms around the vacancy.  Four samples represent a rather
limited set of values for statistical analysis, so the relative
occurrence of $^{13}$C isotopes picked up by these measurements cannot
be used for such analysis.  A much larger number of NV samples is
needed in the spin-echo measurements in order to use the relative
occurrence of the $^{13}$C isotope signals for the identification of
individual atoms in the diamond lattice.

\section{Summary and Conclusions}

In this work we investigated the negatively charged nitrogen-vacancy
center in diamond in detail by \emph{ab initio} supercell calculations using
density functional theory methods.  We showed that the energy sequence
of multiplet states is $^3A_2$, $^1A_1$, $^1E$, $^3E$, $^1E$, and
$^1A_1$. This means that the singlet $^1E$ state enhances the spin
polarization process during the optical cycling of the defect. The
center has non-zero spin ground state. The full hyperfine tensor for a
large number of atoms around the defect was calculated for the first
time. The calculated hyperfine constants of the C ligands agree
well with the experimental values detected by electron paramagnetic
resonance tools. However, there is a controversy about the number of
symmetrically equivalent carbon atoms of the second highest hyperfine
interactions when these are compared to experiment. We propose that that
part of the electron paramagnetic resonance spectrum should be
re-investigated in detail in order to clarify this issue.  Our
calculations reveal that the spin density of the ground state is
spread in the lattice, mostly on a plane perpendicular to the (111)
direction defined by the positions of the N atom and the vacant site,
and that it does not decay monotonically from the vacant site. As a
consequence, only a certain number of $^{13}$C isotopes can interact
with the electron spin, which can be used as qubits for quantum
computing.  Using the limited number of measurements that have been
recently published for single nitrogen-vacancy centers detected by
spin-echo measurements, we were able to identify some individual atoms
around the defect.  Our results contribute to the understanding of the
spin-echo signals of the defect, which is a crucial step towards
realization of the qubit concept in this system.  Additional spin-echo
measurements in NV samples will help identify other
individual $^{13}$C atoms around the defect.

\section*{Acknowledgments}
We are thankful to Liang Jiang, Jeronimo Maze and Mikhail Lukin
for encouraging discussions. 
AG acknowledges support from E\"otv\"os Fellowship of
Hungary. MF acknowledges support by Harvard's Nanoscale Science and
Engineering Center, funded by the National Science Foundation, Award
Number PHY-0117795.

\end{document}